\documentclass[%
twocolumn,
superscriptaddress,
amsmath,
amssymb,
prl,
]{revtex4-2}

\usepackage{graphicx}% Include figure files
\usepackage{dcolumn}% Align table columns on decimal point
\usepackage{physics}
\usepackage{bm}% bold math
\usepackage{braket}
\usepackage{hyperref}% add hypertext capabilities
\usepackage{siunitx}
\hypersetup{
    colorlinks=true,
    linkcolor=blue,
    citecolor=blue,
    urlcolor=blue,
}
\usepackage[capitalise]{cleveref}

\renewcommand{\vec}[1]{\mathbf{#1}}

\newcommand{\qcom}[0]{\vec{Q}_\mathrm{COM}}

\newcommand{\kv}[0]{\vec{K}}   
\newcommand{\qv}[0]{\vec{Q}}   
\newcommand{\mv}[0]{\vec{M}}

\begin{document}

\preprint{APS/123-QED}

\title{Unconventional bright ground-state excitons in monolayer TiI$_2$ from first-principles calculations}

% Force line breaks with \\

\author{Franz Fischer}%
\affiliation{%
Institute of Physical Chemistry, University of Hamburg, Luruper Chaussee 149, 22607 Hamburg, Germany}
\affiliation{%
Max Planck Institute for the Structure and Dynamics of Matter, Luruper Chaussee 149, 22761 Hamburg, Germany
}
\author{Carl Emil Mørch Nielsen}
\affiliation{%
Institute of Physical Chemistry, University of Hamburg, Luruper Chaussee 149, 22607 Hamburg, Germany}

\author{Marta Prada}%
\affiliation{%
Institute of Physical Chemistry, University of Hamburg, Luruper Chaussee 149, 22607 Hamburg, Germany}
\affiliation{I. Institute for Theoretical Physics, University of Hamburg, 22761 Hamburg, Germany}

\author{Gabriel Bester}%
\affiliation{%
Institute of Physical Chemistry, University of Hamburg, Luruper Chaussee 149, 22607 Hamburg, Germany}
\affiliation{The Hamburg Centre for Ultrafast Imaging, Luruper Chaussee 149, 22761
Hamburg, Germany}
\affiliation{\textnormal{corresponding author: \url{gabriel.bester@uni-hamburg.de}}}

\date{\today}% It is always \today, today,

\begin{abstract}
\section{Abstract}
Based on \textit{ab initio} screened configuration interaction calculations we find that TiI$_2$ has a bright exciton ground state and identify two key mechanisms that lead to this unprecedented feature among transition metal dichalcogenides. 
First, the spin-orbit induced conduction band splitting results in optically allowed spin-alignment for electrons and holes across a significant portion of the Brillouin zone around the $\kv$-valley, avoiding band crossings seen in materials like monolayer MoSe$_2$. 
Second, a sufficiently weak exchange interaction ensures that the bright exciton remains energetically below the dark exciton state. 
We further show that the bright exciton ground state is stable under various mechanical strains and that trion states (charged excitons) inherit this bright ground state. 
Our findings are expected to spark further investigation into related materials that bring along the two key features mentioned, as bright ground-state excitons are crucial for applications requiring fast radiative recombination.
\end{abstract}

\maketitle

%\tableofcontents

\section{Introduction}
In semiconductors, photon absorption goes along with the excitation of electrons from the valence to the conduction band, creating electron-hole pairs that interact via Coulomb forces to form excitons \cite{Frenkel31, Wannier37}.
These excitons can recombine radiatively, emitting light -- a process essential for light-emitting diodes \cite{Gu2019_LEDs}, lasers \cite{Wen2020_X_lasers}, and other optoelectronic devices \cite{Mak2016_photonics, Mueller2018_optoelectronic_dev}.
The radiative recombination rate is determined by material properties \cite{Fang2019_X_control_life}, external fields \cite{Mondal2022_photoexcitation}, and temperature \cite{Chen2019_X_life}, and can decrease significantly in two-dimensional (2D) materials due to strong exciton confinement \cite{Palummo2015_X_rad_life}.
Monolayer transition metal dichalcogenides (TMDs), such as MoS$_2$, MoSe$_2$, WS$_2$ and WSe$_2$, and most semiconductor nanostructures, all face a significant challenge: the presence of a dark exciton ground state \cite{Bester03, Deilmann_dark_X_2017, Deilmann_finite_Q_2019, Malic_dark_X, Torche2019_trions, Robert2020_dark_X}, emerging due to the exchange interaction \cite{Bester03,Echeverry_2016_DBS_exchange}. 
Indeed, the electron-hole exchange interaction induces a splitting of the exciton (so-called fine structure) where the energetically low lying exciton state is optically spin-forbidden \cite{Palummo2015_X_rad_life}. 
Exchange also favors the dark state in bulk semiconductors, but due to the lack of confinement and the significantly weaker exchange interaction, compared to the situation in nanostructures, this dark-bright splitting is very small, in the range of $\mu$eV \cite{fu99}, and unproblematic. 
For bulk materials the search for good emitters revolves around finding structures with a direct band gap \cite{fadaly20,yuan24}, rather than a bright ground state.

In nanostructures, and especially in 2D nanostructures with only a few atomic layers, the dark-bright splitting is on the order of room temperature thermal energy, i.e. several tens of meV \cite{Molas_2017_DBS, Ren_2023_DBS_WSe2} and the carrier recombination process is significantly slowed down by the presence of the low-energy dark states, particularly at low temperatures. 
Several approaches have been employed to reduce dark-bright splitting or to brighten the dark state, thereby enhancing the recombination rate. These methods include the application of strain \cite{Chowdhury2024_strain_darkX}, electric fields \cite{Ren_2023_DBS_WSe2}, and magnetic fields \cite{Molas_2017_DBS, Zinkiewicz_2020_brightening_X_mag}. An inversion of bright and dark states has been suggested theoretically based on strong Rashba interactions in halide perovskite quantum dots by Becker et al. \cite{Becker2018}, but was later disproven experimentally by Tamarat et al. \cite{Tamarat2019}. A high-throughput computational search for bright ground state emitters based on Rashba interactions in quantum dots has been performed  recently \cite{Swift2024}, but to the best of our knowledge none of the candidate materials have been experimentally validated to this point.

In this paper, we present \textit{ab initio} calculations on hexagonal monolayer TiI$_2$, a potentially stable two-dimensional transition metal dihalide, demonstrating a bright exciton ground state for two key reasons.
First, the spin-orbit induced conduction band splitting of TiI$_2$ is dominated by the iodide atoms, resulting in parallel spins between conduction band electrons and valence band electrons within an appropriately large radius around the $\kv$-valley favoring a low-energy bright exciton.
Second, the exchange interaction in TiI$_2$ is sufficiently weak, preventing the bright exciton to accumulate repulsive exchange energy, allowing it to remain energetically favorable compared to the dark state.
Our calculations show that the bright exciton ground state remains stable under mechanical strain. The trion states (charged excitons) inherit the exciton properties leading to a bright trion ground state formation. 
Consequently, TiI$_2$ is a promising candidate for fast radiative exciton recombination processes, representing an attractive material for optoelectronic applications. The rather simple underlying physics explaining the bright trion formation could be applied to seek for other bright two-dimensional materials. 

\section{Results}
\begin{figure}
    \centering
    \includegraphics[width=1.0\linewidth]{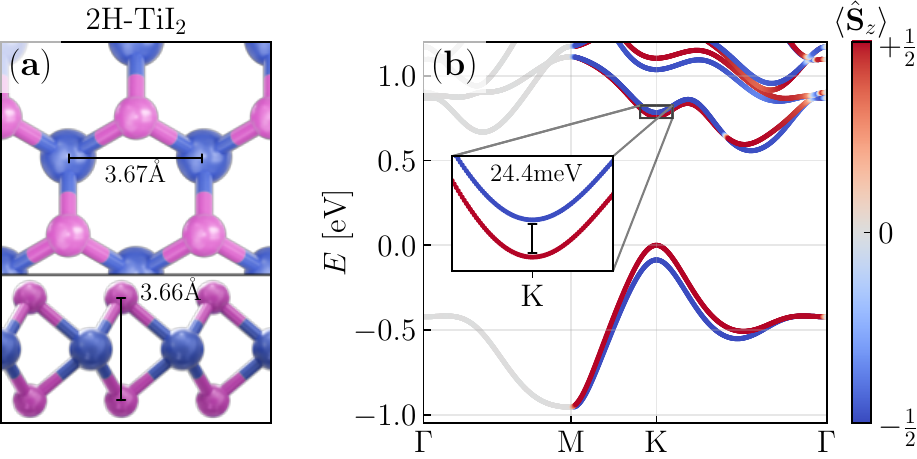}
    \caption{\textbf{Unit cell and band structure of monolayer TiI$_2$.} (\textbf{a}) TiI$_2$ in top and side view with the in-plane lattice constant $a$ and $\mathrm{I}-\mathrm{I}$ distance $d$. (\textbf{b}) The band structure computed within DFT. The color indicates the spin expectation value along $z$. The inset shows the spin-orbit coupling induced conduction band splitting at $\kv$.}
    \label{fig:tii2_dft}
\end{figure}

Previous theoretical predictions showed that TiI$_2$ could prefer the 1T-phase over the 1H-phase depending on the chosen Hubbard $U$ \cite{Li_2020}, making the stability of the 1H-phase an active field of research. Nevertheless, experiments on TMDs have proven that the stabilization of unpreferred polytypes using external stimuli is possible \cite{Li_2021,Han_2025}. We assume monolayer TiI$_2$ to crystallize in the 1H-phase, as this is the semiconducting phase.
We find a DFT-relaxed in-plane lattice constant of $\SI{3.67}{\AA}$ and an iodide-iodide length of $\SI{3.66}{\AA}$, as depicted in \cref{fig:tii2_dft}(a), both roughly $10$-$\SI{15}{\%}$ larger than in usual TMDs \cite{Rasmussen2015}.
We calculated the phonon dispersion relation and observed no imaginary phonon frequencies, suggesting that our monolayer material could be stable \cite{Malyi2019_stability} (see Figure S1).
% To confirm the stability of the system, we calculated the phonon dispersion relation and observed no imaginary phonon frequencies, suggesting that our monolayer material could be stable \cite{Malyi2019_stability} (see Figure S1).
%

At the PBE-DFT level, the electronic structure of TiI$_2$ exhibits an indirect band gap of $\SI{0.56}{eV}$, with the valence band maximum (VBM) located at $\kv$ and the conduction band minimum (CBM) found in the $\qv$-valley, situated mid-way between $\Gamma$ and $\kv$.
Notably, the smallest direct gap, relevant for optical absorption processes that do not require phonon assistance ($\qcom = 0$), is $\SI{0.76}{eV}$ at the $\kv$ point. 
Similar to conventional TMDs the broken inversion symmetry results in spin-orbit induced spin-splittings of the bands, with the most considerable effects observed near the $\kv$-valley. 
The spin-splittings at $\kv$ are $\SI{86}{meV}$ and $-\SI{24}{meV}$ for the valence and conduction band, respectively.
We use the convention $\Delta_{i} = \epsilon_{i}^{\uparrow} - \epsilon_{i}^{\downarrow}$, where $i= \{ \mathrm{v,c} \}$ for the valence and conduction bands, respectively.
A distinct feature of the electronic structure of TiI$_2$, which is not observed in TMDs, is that electrons from the conduction and valence band have parallel spins along the $\mv$ to $\kv$ path. 
In contrast to Mo-based TMDs, the spin-up and spin-down conduction bands do not cross in TiI$_2$, see inset in \cref{fig:tii2_mose2_X_comp}(b).
The orbitals that dominate the band edges are the halogen $p$-orbitals and the titanium $d$-orbitals, similar to regular TMDs.
For the conduction band at $\kv$ the orbitals at play are the in-plane $p$-orbitals and the $d_{z^2}$-orbital (see Figure S2).
The spin-splitting $\Delta_{\mathrm{c}}(\kv)$, induced by spin-orbit coupling (SOC), is mediated to first order by in-plane halogen $p$-orbitals, giving rise to a negative sign. 
The SOC correction at $\kv$ stemming from $d_{z^2}$-orbitals of the titanium atom has a positive sign and appears as a second order correction \cite{Komider2013,Liu2013,Kormnyos2014}. 
We find that in TiI$_2$ the $p$-orbital contribution to the conduction band spin-splitting dominates.
We have computed $\Delta_{\mathrm{c}}(\kv)$ for three other titanium dihalide monolayers (TiF$_2$, TiCl$_2$ and TiBr$_2$) and found that the sign is preserved in all materials, confirming the strong halogen influence on the spin-orbit coupling (see Figure S4).

%Excitons in TiI2
\subsection{Excitons}
\begin{figure}
    \centering
    \includegraphics[width=1.0\linewidth]{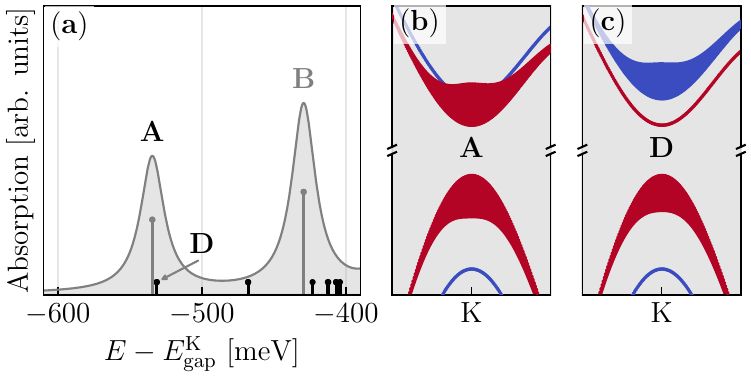}
    \caption{\textbf{Low-energy excitonic states} (\textbf{a}) Energetic positions of many-body optical dipoles are indicated by grey (black) vertical lines for the bright (dark) excitons. (\textbf{b}),(\textbf{c}) Configurational weights at $\kv$ of the first bright (A) and first dark (D) exciton. The color of the bands indicates the spin expectation value along $z$ and the thickness of the bands corresponds to the excitonic weights.}
    \label{fig:tii2_X}
\end{figure}

Since 1H-TiI$_2$ is an indirect semiconductor and the Coulomb interaction is strong throughout the entire Brillouin zone \cite{Deilmann_finite_Q_2019}, excitons with finite center-of-mass momentum ($\qcom \neq 0$) can exist and possess higher binding energies than optically created excitons ($\qcom \approx 0$). In this study we focus on low temperatures ($T < \SI{100}{K}$), where optically generated excitons with $\qcom = 0$ constitute the exciton ground state. This assumption is further supported by our evaluation of the excitonic dispersion, which reveals a local minimum at $\qcom = 0$, see Figure S7.
% The excitons were computed assuming vanishing center-of-mass momentum ($\qcom = 0$) making our results valid for low temperatures.
Furthermore, we have used Kohn-Sham eigenvalues obtained at the PBE-DFT level in the BSE kernel. This is justified, as we are interested in exciton binding energies, i.e. exciton energies relative to the gap.
Additionally, we have computed the $G_0W_0$ corrected band structure of TiI$_2$, validating that the effect of the correction is a nearly rigid shift that has little to no influence on the effective masses (see Supplementary Information and Figure S3).
The low-energy absorption spectrum of TiI$_2$ for the densest $\vec{k}$-grid employed is depicted in \cref{fig:tii2_X}(a) and shows two distinct bright peaks labeled A and B (using the prototypical TMD labeling convention \cite{Mak2010_MoS2, He2014_X_spec}).
% We obtain exciton binding energies of $\SI{441}{meV}$ and $\SI{336}{meV}$ when extrapolating to infinitely dense $\vec{k}$-grids (see Figure S6).
We obtain exciton binding energies of $\SI{441}{meV}$ and $\SI{336}{meV}$ when extrapolating to infinitely dense $\vec{k}$-grids (see Figure S6), with the binding energy of the A exciton being in remarkable agreement with the value of $\SI{440}{meV}$ reported in the computational 2D materials database (C2DB) \cite{c2db_2018}.
The splitting between A and B results mainly from the spin-splittings in the valence and conduction bands and is found to be $\SI{105}{meV}$.
Surprisingly, we find a bright excitonic ground state. 
The dark exciton D is energetically $\SI{3}{meV}$ higher than A. 
At higher levels of theory, i.e. replacing independent-particle (IP) energies in \cref{eq:BSE_matrix_elements} with $G_0W_0$ corrected energies, this state ordering is conserved, but the dark-bright splitting reduces to $\SI{1.5}{meV}$ (see Figure S3). Note that even if the dark-bright splitting vanishes, the emission rate would only be reduced by a factor of two, still qualifying as a novel type of bright emitter.

\Cref{fig:tii2_X}(b) depicts the configurational weights (encoded in the thickness of the bands) of A obtained from squaring the solutions of the exciton many-body Hamiltonian, showing that the A exciton resides in the lowest conduction band and highest valence band around the $\kv$ point, and both bands have the same spin projection (both red bands in \cref{fig:tii2_X}(c)). 
D is dark, because it is made of configurations that involve electrons and holes of opposite spins (blue and red bands in \cref{fig:tii2_X}(c)) leading to spin-forbidden optical transitions.
Furthermore, we find that the configurational weights of A are slightly larger at $\kv$ ($\SI{4.6}{\%}$) compared to D ($\SI{3.0}{\%}$), whose weights are slightly more spread in $\vec{k}$-space around the $\kv$ point.
For both states, the configurational weights are quite delocalized in reciprocal space, demanding a dense $\vec{k}$-grid sampling to achieve convergence (see Figure S6).

%Trions in TiI2
\subsection{Trions}
\begin{figure}
    \centering
     \includegraphics[width=1.0\linewidth]{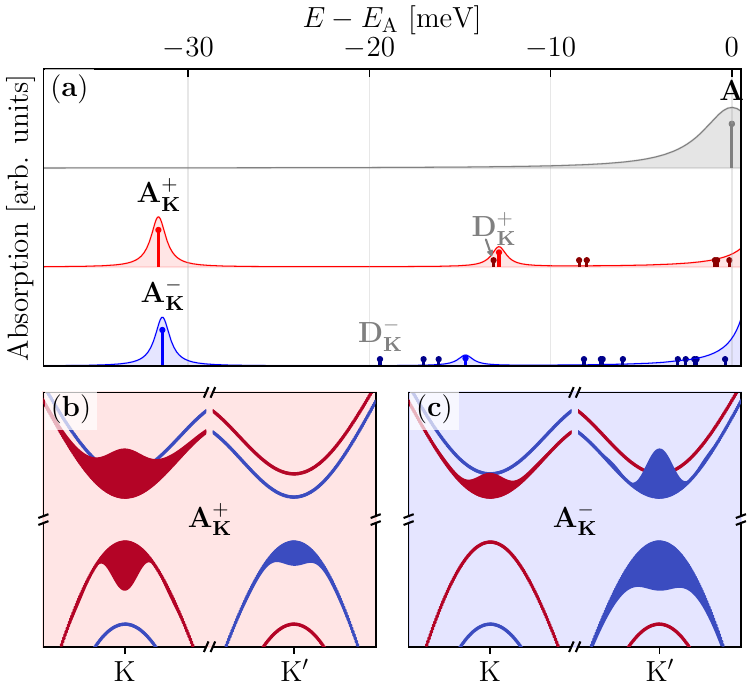}
    \caption{\textbf{Low-energy trion states} (\textbf{a}) Absorption spectrum of positive (red) and negative (blue) trions relative to the bright ground-state exciton A (grey). Energetic positions of many-body optical dipoles are indicated by vertical lines. The dark states, characterized by negligible dipole moments, are not visible at this scale. However, we use small vertical dark lines to indicate their energetic positions. (\textbf{b}),(\textbf{c}) Configurational weights at $\kv$ of the first bright positively (A$^{+}_{\kv}$) and negatively (A$^{-}_{\kv}$) charged trion species. The color of the bands indicates the spin expectation value along $z$ and the thickness of the lines encodes the configurational weight.}
    \label{fig:tii2_xmk_xpk}
\end{figure}

In the following section, we focus on trions -- charged excitons that frequently occur in TMDs due to their high binding energies and the presence of excess carriers \cite{Liu2019_WSe2_spectrum_assume_K, Lyons2019_WSe2_spectrum_assume_K, Calman2020}. 
These quasiparticles significantly influence optoelectronics, offering opportunities for manipulating light-matter interactions at the nanoscale, with potential applications in quantum information processing and photonic devices. 
The screened CI methodology \cref{eq:eff_mb_Hamiltonian} can be naturally extended to trions \cite{MørchNielsen2025}, however the center-of-mass momentum $\qcom$ of the three-particle compound has to be chosen. 
We choose the $\kv$ point ($\qcom = \kv$), meaning that the additional electron, for negatively charged trions, occupies the lowest conduction band at $\kv$, while the additional hole, for positively charged trions, occupies the VBM at $-\kv$ (or $\kv^{\prime}$).
Notably, the conduction band at the $\qv$ valley (between the $\kv$ and $\Gamma$ points) is lower in energy than at $\kv$. 
Previous studies on monolayer WSe$_2$ also revealed that negatively charged trions involving the $\qv$ valley form at lower energies and would thus be energetically more favorable \cite{MørchNielsen2025}.
However, as the lowest energy excitons form around $\kv$, we have limited our investigation to the $\kv$ valley.
The absorption spectra for the trions are shown in \cref{fig:tii2_xmk_xpk}(a), revealing a bright trion ground state that is energetically nearly independent of charge.
We find trion binding energies for both bright trion states (A$^{-}_{\kv}$ and A$^{+}_{\kv}$) to be $\SI{32}{meV}$ with respect to A. 
The configurational weights of A$^{-}_{\kv}$ and A$^{+}_{\kv}$ shown in \cref{fig:tii2_xmk_xpk}(b),(c) indicate that the trion states are inter-valley singlet states resembling A in one valley with an additional carrier of the opposite spin in the opposite valley.
By a comparison of the weights in \cref{fig:tii2_X}(b) and \cref{fig:tii2_xmk_xpk}(b),(c) we notice that the trion configurations are more localized in reciprocal space compared to the corresponding exciton.
Especially the additional ``spectator'' carrier (electron at $\kv$ in A$^{-}_{\kv}$ and hole at $\kv '$ in A$^{+}_{\kv}$) is more localized in reciprocal space and hence more spread out in real space.
In contrast to the binding energy, the dark bright splitting is charge-sensitive, namely $\SI{18}{meV}$ for A$^{+}_{\kv}$ and $\SI{12}{meV}$ for A$^{-}_{\kv}$.
Convergence results for both trion species can be found in Figure S8.

\subsection{Strain effects}
\begin{figure}
    \centering
    \includegraphics[width=1.0\linewidth]{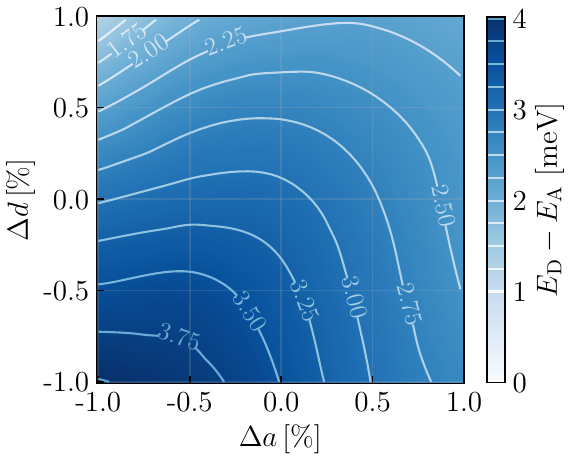}
    \caption{\textbf{Dark-bright splitting under the influence of strain} The in-plane lattice constant $a$ and the I-I distance $d$ have been varied within the interval $[-1, +1] \, \%$ from their equilibrium values.}
    \label{fig:tii2_strain}
\end{figure}

Lastly, we have tested the robustness of the bright exciton ground state in monolayer TiI$_2$ by varying the in-plane lattice constant $a$ and the I-I distance $d$, see \cref{fig:tii2_strain}. 
Remarkably, across all tested strains (ranging from $-1\%$ to $+1\%$ from equilibrium), the exciton ground state remains bright. 
In an experimental setting, where the structure is compressed in-plane (negative $\Delta a$) the structure would typically expand out-of-plane (positive $\Delta d$) so that one would move close to a diagonal line in \cref{fig:tii2_strain}. 
Starting from the relaxed structure ($\Delta a = \Delta d = 0$) a compressive strain (smaller $a$, larger $d$) or a tensile strain (larger $a$, smaller $d$) would both lead to a decrease in the dark-bright splitting, with a more rapid reduction observed under compressive strain.
The data indicates that the dark-bright splitting may reverse sign under sufficiently large compressive strains, providing an experimentally accessible tuning knob between a dark and bright exciton ground state \cite{Frisenda2017_strain_tmd, Chowdhury2024_strain_darkX}. We also found that the dark-bright splitting is enhanced for smaller $a$ and $d$, but applying such a strain may be experimentally challenging.

\section{Discussion}
\begin{figure}
    \centering
    \includegraphics[width=1.0\linewidth]{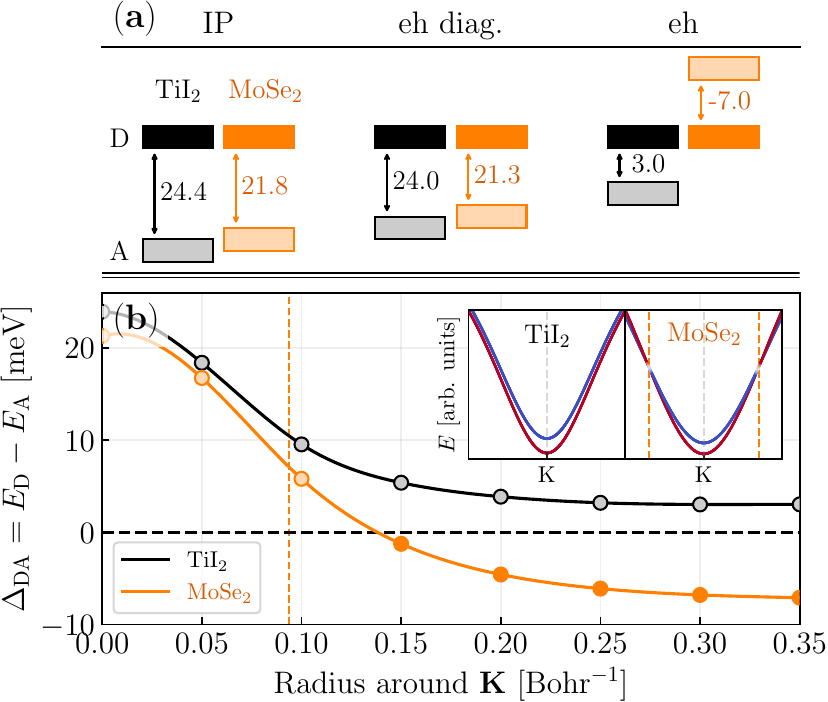}
    \caption{\textbf{Dark-bright splitting analysis} (\textbf{a}) Energetic splitting in meV between the dark (D) and the bright state (A) in monolayer TiI$_2$ (black/grey) and MoSe$_2$ (light-orange/orange) calculated at different levels in the exciton calculations: using only single-particle energies (IP), including diagonal electron-hole interactions (eh diag.) and including all interactions (eh) in \cref{eq:BSE_matrix_elements}. (\textbf{b}) Dark-bright splitting ($\Delta_{\mathrm{DA}}$) for TiI$_2$ and MoSe$_2$ as a function of the number of configurations (given by the radius around $\kv$). The inset shows the conduction bands for both materials around $\kv$ and the orange vertical lines indicate the radius $r_{\times}$ for which the spin-split conduction bands cross in MoSe$_2$.}
    \label{fig:tii2_mose2_X_comp}
\end{figure}

In the subsequent analysis of the exciton dark-bright splitting, we define $\Delta_{\mathrm{DA}} = E_{\mathrm{D}} - E_{\mathrm{A}}$ and show the results qualitatively, including the splittings in meV, for monolayer TiI$_2$ and monolayer MoSe$_2$ in \cref{fig:tii2_mose2_X_comp}(a) using different levels of theory.
The sign of $\Delta_{\mathrm{DA}}$ determines whether the excitonic ground state is dark ($\Delta_{\mathrm{DA}} < 0$) or bright ($\Delta_{\mathrm{DA}} > 0$).
When only single-particle energies (IP) are included, there are no interactions between configurations and thus the difference between the bright and dark exciton is, by definition, the negative conduction band splitting, i.e. $\Delta_{\mathrm{DA}} = -\Delta_{\mathrm{c}}(\kv) > 0$. 
We chose monolayer MoSe$_2$ to compare with TiI$_2$, as both materials posses a similar conduction band spin splitting at $\kv$ and hence similar $\Delta_{\mathrm{DA}}$ at the IP theory level.
Next, we allow direct and exchange Coulomb integrals on the diagonal of the BSE matrix (eh diag.), reducing $\Delta_{\mathrm{DA}}$ by $\SI{0.4}{meV}$ and $\SI{0.5}{meV}$ for TiI$_2$ and MoSe$_2$, respectively.
Both excitonic states are equally shifted down in energy by the attractive direct Coulomb interaction (not shown since we align the D-state energetically in \cref{fig:tii2_mose2_X_comp}(a)) but A gains a small energy from the repulsive exchange interaction, while exciton D does not, due to the improper spin alignment. For TiI$_2$ (MoSe$_2$) this leads to the reduction of the splitting from $\SI{24.4}{meV}$ ($\SI{21.8}{meV}$) to $\SI{24.0}{meV}$ ($\SI{21.3}{meV}$).
Notably, the diagonal exchange interaction is slightly larger in MoSe$_2$ than in TiI$2$ ($\SI{0.5}{meV}$ vs. $\SI{0.4}{meV}$) at this low level of the theory.
This effect becomes amplified when the coupling between configurations is allowed via off-diagonal Coulomb interactions (eh, our final result). 
In this case, the bright state A is shifted up by $\SI{21}{meV}$ in TiI$_2$ and by $\SI{28.3}{meV}$ MoSe$_2$, leading to $\Delta_{\mathrm{DA}} = \SI{3}{meV}$ for TiI$_2$ and $\Delta_{\mathrm{DA}} = -\SI{7}{meV}$ for MoSe$_2$.
Here, the exchange coupling in MoSe$_2$ is so strong that it reverses the sign of $\Delta_{\mathrm{DA}}$ by pushing A higher in energy than D, resulting in a dark exciton ground state for MoSe$_2$, as in most nanostructures.
Previous theoretical studies have predicted a value of $\Delta_{\mathrm{DA}} = -\SI{10}{meV}$ for MoSe$_2$, which is in excellent agreement with our findings \cite{Deilmann_dark_X_2017, Deilmann_finite_Q_2019}.

% Dispersion Idea
In addition to the weaker exchange interaction in TiI$_2$, the dispersion of the conduction bands around $\kv$ influences whether the exciton ground state is bright or dark. 
To illustrate this effect, we plot in \cref{fig:tii2_mose2_X_comp}(b) $\Delta_{\mathrm{DA}}$ as a function of the radius around $\kv$, the latter determining the number of configurations taken into account in \cref{eq:BSE_EVP}, see also Figure S5.
A vanishing radius around $\kv$ is equivalent to the IP result $\Delta_{\mathrm{DA}} = - \Delta_{\mathrm{c}}(\kv)$.
In MoSe$_2$, the conduction bands cross at a radius of $ r_{\times} \sim \SI{0.09}{Bohr^{-1}} $ from $\kv$ towards $\mv$, as shown in the right inset of \cref{fig:tii2_mose2_X_comp}(b) where $r_{\times}$ is marked with a dashed vertical line.
For configurations when the radius in reciprocal space is larger than $r_{\times}$ in MoSe$_2$, the conduction band ordering (blue below red in the right inset of \cref{fig:tii2_mose2_X_comp}(b)) leads to a negative $\Delta_{\mathrm{DA}}$ at the IP level of theory due to the band crossings. In contrast, TiI$_2$ exhibits no such band crossings, resulting in a positive $\Delta_{\mathrm{DA}}$ contribution across the entire reciprocal space around the $\kv$-point.
This effect can be seen in the main panel of \cref{fig:tii2_mose2_X_comp}(b) where $\Delta_{\mathrm{DA}}$ drops below zero (dark exciton ground state) for 
radii $ \gtrsim \SI{0.14}{Bohr^{-1}} $ in MoSe$_2$, as more of these energetically unfavorable configurations are added to A.
For TiI$_2$ this effect is absent, as the conduction bands do not cross and $\Delta_{\mathrm{DA}}$ remains positive. 
Moreover, none of the other titanium dihalide monolayers (TiF$_2$, TiCl$_2$, and TiBr$_2$) exhibit a bright ground state exciton, and all display a crossing of the conduction bands similar to that observed in MoSe$_2$, highlighting the significance of this absence of band crossing.

%\section{Conclusion}
To summarize, we have identified a hexagonal monolayer material that exhibits a bright exciton ground state. 
We single out two reasons that allow this yet unobserved feature in transition metal dichalcogenides.
First, the highest occupied valence and lowest unoccupied conduction bands show the same spin-expectation value for a large portion of the Brillouin zone around the $\kv$-valley and do not cross, unlike in MoSe$_2$, where spin-orbit coupling induces band crossing.
Second, while the exchange interaction favors the dark exciton, its contribution in TiI$_2$ is insufficient to energetically shift the bright state above the dark state.
Additional calculations of positive and negative trions reveal that their ground state remains bright regardless of charge, mirroring the behavior of neutral excitons.
Finally, we have applied strain to our system, demonstrating that the bright exciton ground state is robust and that the dark-bright splitting can be experimentally tuned through mechanical deformation.
We believe that our discoveries hold significant implications for experimental exploration of monolayer TiI$_2$, as its distinctive optical properties have the potential to facilitate rapid recombination processes, which are essential for the advancement of future light-emitting diodes, lasers, optoelectronic devices, and quantum technology. 
Our derived understanding of the underlying process also provides a framework for identifying other materials with similar properties.
However, since many-body effects driven by Coulomb interactions appear to be the leading mechanism, establishing a universal filtering criterion for such materials remains an open challenge.

\section{Methods}

\subsection{Structural and electronic properties}
Throughout this work, the electronic ground state is found from density functional theory (DFT) calculations using the \texttt{Quantum Espresso} suite \cite{Gianozzi2009, Gianozzi2017}, employing the Perdew-Burke-Ernzerhof (PBE) exchange-correlation functional \cite{Perdew1996}.
We utilize fully-relativistic, norm-conserving pseudopotentials from the \texttt{PseudoDojo} library \cite{vanSetten2018}.
We optimize atomic positions until forces fall below $10^{-4}$ Ry/Bohr, with the out-of-plane cell length fixed to 45 Bohr to ensure adequate vacuum and eliminate spurious interactions between periodic images. 
During the relaxation step we use an increased plane-wave cutoff energy of $\SI{90}{Ry}$, while for all other calculations, a cutoff of $\SI{80}{Ry}$ provided converged results. 
Additionally, we computed phonons within density functional perturbation theory using an energetic cutoff of $\SI{140}{Ry}$ on a $24 \times 24 \times 1$ $\vec{k}$-point, and a $9 \times 9 \times 1$ $\vec{q}$-point Monkhorst grid, while truncating the Coulomb interaction to two dimensions \cite{Sohier_assume_2D}. 

\subsection{Many-body excited states}
We apply an \textit{ab initio} effective many-body Hamiltonian, $\mathcal{H}$, formulated in second quantization using the canonical transformation to the electron-hole picture. 
This screened configuration interaction (CI) approach has proven reliable for semiconductor quantum dots \cite{Franceschetti1999,Bester_2009} and 2D materials \cite{Torche2019_trions, MørchNielsen2025}. 
The Hamiltonian is expressed as
\begin{equation}
    \mathcal{H} = \mathcal{H}_0 + \mathcal{H}_{eh} + \mathcal{H}_{ee} + \mathcal{H}_{hh},
    \label{eq:eff_mb_Hamiltonian}
\end{equation}
where $\mathcal{H}_0$ corresponds to independent-particle energies, while $\mathcal{H}_{eh}$, $\mathcal{H}_{ee}$ and $\mathcal{H}_{hh}$ account for electron-hole, electron-electron and hole-hole interactions. 
These interactions are Coulombian and thus related to two-body matrix elements computed from wave functions obtained through DFT.
Projecting \cref{eq:eff_mb_Hamiltonian} into the exciton subspace yields the well-known Bethe-Salpeter equation (BSE) in the Tamm-Dancoff approximation \cite{Onida2002_bse}.
The matrix element representation of the BSE, using Roman indices for electrons and Greek indices for holes, is given by
\begin{equation}
    \braket{i\alpha|H^{\mathrm{BSE}}|j\beta} = \underbrace{(\epsilon_i - \epsilon_{\alpha}) \delta_{ij} \delta_{\alpha \beta}}_{\text{IP}} -\underbrace{\braket{i \beta|W|j \alpha}}_{\text{direct}} + \underbrace{\braket{i \beta|v|\alpha j}}_{\text{exchange}}.
    \label{eq:BSE_matrix_elements}
\end{equation}
The IP energies are obtained either from DFT calculations or from single-shot $G_0W_0$ calculations within the quasiparticle approximation, as indicated in the main body of the text.
The $G_0W_0$ calculations are performed using \texttt{YAMBO}, employing the plasmon pole approximation for the dynamical dielectric screening.
The direct Coulomb matrix elements incorporate the screened interaction kernel, $W$, with static dielectric screening computed within the random phase approximation using the \texttt{YAMBO} code \cite{Marini2009_yambo, Sangalli2019}.
The electron-hole exchange interaction is treated unscreened ($v$), as derived from the exchange part of the electron self-energy \cite{Onida2002_bse, Torche2021}.
We can reformulate \cref{eq:BSE_matrix_elements} to an eigenvalue problem:
\begin{equation}
    \sum_{j\beta} \braket{i\alpha|H^{\mathrm{BSE}}|j\beta} A^{\lambda}_{j\beta}= E_{\lambda} A_{i\alpha}^{\lambda},
    \label{eq:BSE_EVP}
\end{equation}
yielding the exciton energies, $E_{\lambda}$, and the complex exciton wave function coefficients, $A_{i\alpha}^{\lambda}$, whose absolute square corresponds to the configurational weights of state $\lambda$.
For trions, an eigenvalue problem can be similarly formulated by projecting \cref{eq:eff_mb_Hamiltonian} into the three-particle subspace of the respective trion species. A more sophisticated and detailed formulation is available in \cite{MørchNielsen2025}.
Computationally relevant parameters and convergence tests are given in the Supplementary Information.

\section{References}
\bibliography{bibliography}

\begin{acknowledgments}

\section{Data availability}
The data used and analyzed during the current study is available from the corresponding author upon reasonable request.

\section{Code availability}
The underlying code for this study is not publicly available but may be made available upon reasonable request from the corresponding author.

\section{Acknowledgements}
The project is supported by the Deutsche Forschungsgemeinschaft (DFG, German Research Foundation) within the Priority Program SPP2244 2DMP and by the Cluster of Excellence ``Advanced Imaging of Matter'' of the DFG -- EXC 2056 -- project ID 390715994. 
Calculations were carried out on Hummel funded by the DFG – 498394658.
\end{acknowledgments}

\section{Author contributions}
F.F. conceptualized the research, conducted all calculations, analyzed and visualized the results. Method development was carried out by F.F. and C.E.M.N.. All authors (F.F., C.E.M.N., M.P. and G.B.) contributed equally to the writing of the manuscript.

\section{Competing interests}
The authors declare no competing interests. 

\end{document}